\documentclass[runningheads]{llncs}
\usepackage[T1]{fontenc}
\usepackage{graphicx}
\usepackage{amsmath,amssymb,amsfonts}
\usepackage{algorithmic}
\usepackage{textcomp}
\usepackage{listings}
\usepackage{pgfplots}
\usepackage{pgfplotstable}
\usepackage{subcaption}
\usepackage{csquotes}
\usepackage{multirow}
\usepackage{colortbl}
\usepackage{tabularx}
\usepackage{booktabs}
\usepackage{hyperref}
\usepackage[capitalise,noabbrev]{cleveref}
\usepackage{xcolor}
\usepackage[inline,shortlabels]{enumitem}
\usepackage[scaled=.98]{inconsolata}

\usetikzlibrary{patterns, patterns.meta}
\usepgfplotslibrary{groupplots}

\pgfplotsset{compat=1.18}

\newenvironment{enumi}{\begin{enumerate*}[label=(\roman*)]}{\end{enumerate*}}

\newcommand{\NHRText}{%
The authors would like to thank the Federal Ministry of Research, Technology and Space
and the state governments (www.nhr-verein.de/unsere-partner) for supporting this
work as part of the joint funding of National High Performance Computing
(NHR).
}

\begin{document}
\title{Allocation Tracking and Parameter Checking for Parallel~Programming~Models using Contracts}
\titlerunning{Allocation Tracking and Parameter Checking using Contracts}
%
\author{Yussur Mustafa Oraji\inst{1}\orcidID{0009-0004-9922-3112} \and
Christian~Bischof\inst{1}\orcidID{0000-0003-2711-3032}}
\authorrunning{Y. Oraji et al.}
%
\institute{Department of Computer Science, TU Darmstadt University, Darmstadt, Germany \email{\{yussur.oraji,christian.bischof\}@tu-darmstadt.de}}
\maketitle              
\begin{abstract}
Correctness checking tools for High-Performance Computing programs are typically limited to specific parallel programming models such as MPI or OpenSHMEM.
The CoVer framework previously addressed this by introducing a generic, contract-based approach that decoupled API requirements from the core tool.
However, CoVer's effectiveness remains bounded by the expressiveness of its underlying contract language, restricting the types of errors it can verify.
This paper presents an extension to the CoVer contract language designed to capture and check a broader range of error classes.
Our extensions introduce generic parameter checking and allocation tracking,
while keeping generality across both programming model and language.
We evaluate these extensions and demonstrate that analysis accuracy remains consistent across multiple languages,
reinforcing the framework's general applicability.
While the additional runtime analyses naturally incur a performance overhead,
these improvements greatly enhance CoVer's utility with a significant accuracy improvement.

\end{abstract}

\section{Introduction}
\label{sec:introduction}

Parallel programming models for distributed-memory computing such as MPI \cite{messagepassinginterfaceforumMPIMessagePassingInterface2025} and OpenSHMEM \cite{openshmemcommitteeOpenSHMEMApplicationProgramming2020}
allow for efficient, large-scale computations on high-performance clusters.
As they are quite low-level, the responsibility of correct use falls on the programmer,
requiring them to manually ensure correct synchronization, memory management, or use of API constants.
Errors manifest as crashes, deadlocks, memory corruption or silently as incorrect results.

Dynamic correctness checkers such as MUST/RMASanitizer \cite{hilbrichMUSTScalableApproach2010,schwitanskiRMASanitizerGeneralizedRuntime2024} and PARCOACH \cite{saillardPARCOACHCombiningStatic2014} allow for the automatic detection of programming errors at runtime.
They exhibit high accuracy, but cause significant runtime overhead, and can only check materialized errors but not all potential ones.
In contrast, static correctness checkers such as the SPMD IR \cite{burakSPMDIRUnifying2025} and MPI-Checker \cite{drosteMPIcheckerStaticAnalysis2015} perform detection at compile time,
which allows for minimal overhead, and allow reporting of all possible errors due to their whole program view. 
However, they instead have poor accuracy due to their inability to inspect runtime context.
Furthermore, all tools listed support only a fixed set of programming models and API functions therein, making it difficult to extend detection for further error classes or API functions, requiring the modification of the tool itself.

In previous work, we presented CoVer \cite{orajiVerifyingMPIAPI2026}, a static, contract-based correctness checker for parallel programming models.
By using contracts, which can be attached to arbitrary API functions or programming models even by users, CoVer allows for a high degree of modularity.
Each contract may contain pre- and postconditions, which describe the requirements before and after each callsite respectively.
We later introduced CoVer-Dynamic \cite{orajiDynamicContractAnalysis2026}, which can analyze the same contracts at runtime to allow for a higher degree of detection accuracy.

The dependence on the contract language does have a downside: While arbitrary error classes are supported in theory, they must be representable using the contract language.
\begin{figure}[tbp]
    \begin{subfigure}{0.45\textwidth}
        \begin{lstlisting}
int* buf = (int*)malloc(...);
if (rank == 0) {
  // Parameter error
  MPI_Send(buf, ..., ^*\textbf{MPI\_ANY\_SOURCE}*^);
} else if (rank == 1) {
  // Use after free
  ^*\textbf{free(buf);}*^
  MPI_Recv(^*\textbf{buf}*^, ..., ^*MPI\_ANY\_SOURCE*^);
}
        \end{lstlisting}
        \caption{Parameter and allocation errors}
        \label{fig:simple_err_example}
    \end{subfigure}
    \hfill
    \begin{subfigure}{0.45\textwidth}
        \begin{lstlisting}
int MPI_Send(...) CONTRACT(
  PRE {param!(3:!=MPI_ANY_SOURCE)}
);

int MPI_Recv(...) CONTRACT(
  PRE {alloc!(0)}
);
        \end{lstlisting}
        \caption{Violated Requirements / Contracts}
        \label{fig:simple_err_contracts}
    \end{subfigure}
    \caption{Examples of parameter and allocation errors. The first contract requires that the destination rank is not \texttt{MPI\_ANY\_SOURCE}, and the second that the buffer must be allocated.}
\end{figure}
Some errors such as common invalid parameter issues or memory use-after-free are currently not representable,
such as those in \cref{fig:simple_err_example}.

This paper presents an extension to the CoVer contract language for parameter checking and allocation tracking,
allowing for a higher degree of freedom when interacting with the CoVer verification framework.
The extension allows describing the aforementioned errors using an expanded contract syntax,
exemplified in \cref{fig:simple_err_contracts}.
We develop both static data-flow and runtime analyses for these newly definable contracts,
and implement support for both the original C and the new Fortran port of CoVer \cite{orajiExtendingContractVerification2026}.

\section{Background}
\label{sec:background}

Our goal is the implementation of parameter checking and allocation tracking to the CoVer verification framework.
While both the static and dynamic analysis of CoVer, hereafter referred to as CoVer-Static and CoVer-Dynamic, utilize the same contract language,
they work in entirely different ways: the former using a formal data-flow at compile time, the latter with runtime callbacks.
This section introduces the contract language as well as the different analysis modes in detail.

\subsection{CoVer Contract Language}
\label{sec:background_contract_language}

The CoVer contract language is built using preconditions and postconditions, which are attached to API functions.
\begin{figure}[tbp]
    \begin{subfigure}{0.45\textwidth}
        \begin{lstlisting}
int MPI_Isend(^*\dots*^) CONTRACT(
  PRE { call!(MPI_Init) }
  POST {
    no! (write!(*0))
    until! (call!(MPI_Wait(0:6)))
  }  
);
        \end{lstlisting}
        \caption{Defining a contract in C.}
    \end{subfigure}
    \hfill
    \begin{subfigure}{0.45\textwidth}
        \begin{lstlisting}[language=fortran]
call Declare_Contract(MPI_Isend, &
  "PRE { call!(MPI_Init) } &
 & POST { &
 &  no! (write!(0)) &
 &  until!(call!(MPI_Wait(0:6))) &
 & }"
);
        \end{lstlisting}
        \caption{Defining a contract in Fortran.}
    \end{subfigure}
    \caption{Simple examples of CoVer contracts.}
    \label{fig:simple_contract}
\end{figure}
This can be seen in \cref{fig:simple_contract}, which demonstrates two requirements.

First, it states that before performing a communication using \texttt{MPI\_Isend} the MPI library must be initialized.
The example contract contains an \emph{operation}, more specifically a call operation,
which is placed in the precondition.
Thus, the contract states that before a call to the contract supplier (the function the contract is attached to, here: \texttt{MPI\_Isend}),
the function contained in the call operation (here: \texttt{MPI\_Init}) must already have been called.

Further, the contract requires that after calling \texttt{MPI\_Isend} no writes may occur on the buffer until \texttt{MPI\_Wait} is called using the same request handle,
which avoids data races as \texttt{MPI\_Isend} is a non-blocking communication call.
Instead of a call operation this requirement makes use of a \emph{release} operation,
which is used to specify that an operation \emph{may not occur until} another operation releases the requirement.
In this case, the forbidden operation is a writing access to the first parameter (indicated by the \texttt{0} in the write operation), which houses the buffer of \texttt{MPI\_Isend}.
The releasing operation is a call to \texttt{MPI\_Wait} where the first parameter of \texttt{MPI\_Wait} must match the seventh of \texttt{MPI\_Isend}, which contain the corresponding request handles.

The example contracts given are simplifications, neglecting e.g. the MPI sessions model, threading, or different communication synchronization mechanisms.
In general, the operations are specified in formulas given in conjunctions,
disjunctions or exclusive-or relations to create more complicated rulesets.
However, for the purposes of this paper this level of knowledge is sufficient.

The limiting factor for parameter checking and allocation tracking are the operations:
Currently it is not possible to compare values to other known ones, checking for null input or whether an input pointer is allocated or not.
Thus, the main factor for the contract language extension is adding the necessary operations to track these values.

\subsection{CoVer-Static}
\label{sec:background_static}

CoVer-Static is built on top of the LLVM compiler infrastructure \cite{LLVMCompilerInfrastructure},
and operates on the level of the LLVM Intermediate Representation (LLVM IR).
\begin{figure}[tbp]
  \centering
  \includegraphics[width=0.8\textwidth]{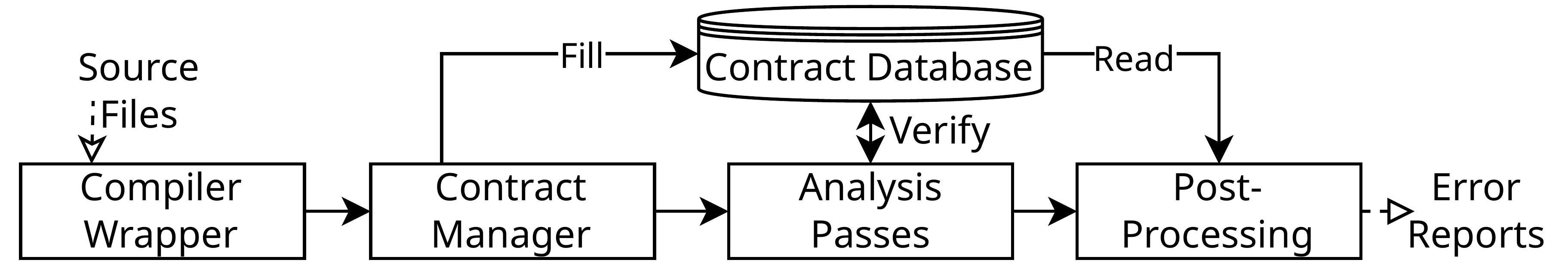}
  \caption{Overview of the CoVer-Static architecture, from \cite{orajiVerifyingMPIAPI2026}}
  \label{fig:static_architecture}
\end{figure}
It is split into four parts, which can be seen in \cref{fig:static_architecture}:
The compiler wrapper, contract manager, the analyses passes and a post-processing step.

The compiler wrapper allows for whole-program analysis by intercepting the actual compilation steps.
Instead of outputting object files, it instead generates LLVM IR for each source file.
When the wrapper detects a link step, it merges all IR files together, and only then starts the analysis.

The analysis passes rely on the contract manager to parse the annotations.
It creates the contract database, which collects all requirements to be verified.

The analysis passes are responsible for checking the fulfillment of the requirements stored in the database.
Each pass is designed to handle a specific type of requirement: \texttt{PreCall}, \texttt{PostCall} and \texttt{Release}.
For example, the contract given in \cref{fig:simple_contract} contains one \texttt{PreCall} requirement and one \texttt{Release} requirement.
The relevant pass performs a formal data-flow analysis, which will be elaborated on in \cref{sec:static_alloc_tracking}.
Finally, it writes the result of its analyses back into the database.

The last step of CoVer-Static is post-processing, in which contracts given in more complicated formulas (conjunctions, disjunctions, exclusive-or) are resolved,
and the error output is performed.

For our extension, two parts of this pipeline must be modified.
First, since we are extending the contract language by adding new operations (as mentioned in \cref{sec:background_contract_language}),
the contract manager must be extended to support parsing the new operations.
Additionally, new analysis passes must be written to support the corresponding requirements.

\subsection{CoVer-Dynamic}
\label{sec:background_dynamic}

In contrast to the compile-time tool CoVer-Static, CoVer-Dynamic performs analysis at runtime using a callback system.
It defines runtime analogues of the analysis passes of CoVer-Static, which each are responsible for the same kinds of requirements as CoVer-Static.
But, CoVer-Dynamic differs in a few key aspects due to necessary changes for runtime analysis.

First, the contract annotations are no longer available at runtime.
For C they are added as function attributes, and for Fortran they are added as calls to an unreachable dummy function.
Thus, the first step to performing the correctness analysis is providing access to the contract definitions.

This is done by adding an initialization callback to CoVer-Dynamic at the start of the program.
The callback is added using another compile-time pass, similarly to those used for CoVer-Static,
but instead of performing verification it instead modifies the code to add the necessary callbacks for CoVer-Dynamic.
In particular, the initialization callback receives a pointer to the contract database, which can then be read at runtime.

Additional callbacks are required for the actual analysis,
more specifically callbacks are added for every function call where the called function is referred to in a contract or has a contract attached.
Further, memory accesses are instrumented as well to allow for detection of, e.g., data races using the contract given in \cref{fig:simple_contract}.
The callbacks are inserted just before the relevant memory access or function call.
This allows for error reporting just before the actual erroneous action in case the action leads to a crash or deadlock.

These callbacks allow the runtime analyses to gather the necessary information for error reporting.
Due to the contract-based nature of CoVer using existing interfaces such as the commonly used PMPI is not possible.
This would create a dependency on MPI, thus contradicting the generic nature of CoVer.

Another difference compared to CoVer-Static is the post-processing step, or the lack thereof in CoVer-Dynamic.
Since buggy programs are likely to crash or deadlock it is not possible to use a post-processing step in CoVer-Dynamic to resolve the contract formulas.
Instead, we use a special \emph{state verification} step, which is performed any time an analysis is either fulfilled or violated.
The state verification checks, given the current state of each contract, whether the full formula is violated and performs
error output if necessary.

To support the parameter error analysis, CoVer-Dynamic mostly requires just modifying the initialization routine in order to generate the needed analyses.
However, major changes are necessary specifically for allocation tracking, as the current callback system does not deliver enough information for it to work.
For example, given a callback for a call to \texttt{MPI\_Isend}, CoVer-Dynamic has no way of knowing whether the given pointer used as the buffer is allocated or not.
The pointer may be a stack variable, a global variable, it may point to heap memory allocated using \texttt{malloc} or similar calls,
or simply be null or pointing to random memory as it has not been initialized.
Even restricting ourselves to just heap memory, adding a callback for \texttt{malloc} would not be sufficient.
Since the callbacks are always added just before the actual function call, we do not have access to the return value.
Thus, the callback system must be reworked entirely, and a method must be devised to infer both stack and global variables.

\section{Implementation}
\label{sec:implementation}

CoVer is composed of three main parts,
\begin{enumi}
    \item the contract language,
    \item the static analysis and
    \item the dynamic analysis.
\end{enumi}
Each of these must be extended to support the additional error classes.

\subsection{Contract Language}
\label{sec:impl_language}

The contract language is mostly limited by the operations;
the current call and release operations cannot describe the parameter or allocation errors.
Thus, we introduce three new contract operations, one for parameter and two for allocation errors respectively.

\begin{figure}[tbp]
    \begin{lstlisting}
// Allocators
void* calloc(size_t num, size_t size) CONTRACT ( POST {alloc!(R[0 * 1])} );
int MPI_Win_allocate(...) CONTRACT ( POST {alloc!(*4[0])} );
// Deallocators
void free(void* ptr) CONTRACT( POST {free!(0)} );
    \end{lstlisting}
    \caption{Examples of allocation/deallocation functions}
    \label{fig:allocator_funcs}
\end{figure}
\paragraph{Allocation Tracking}
The language extension for allocation tracking is the simplest, consisting of new \emph{free} and \emph{alloc} operations.
Requiring a pointer to be allocated is done by simply using the alloc operation in the precondition,
as done in \cref{fig:simple_err_contracts} for the first parameter (the buffer) of \texttt{MPI\_Recv}.

In order for CoVer to recognize which functions allocate or free memory, they must have contracts attached as well.
This can be seen in \cref{fig:allocator_funcs} for a selection of C library functions and \texttt{MPI\_Win\_allocate}.
Determining a function as an allocator is done by using the \texttt{alloc} operation in the \texttt{POST} scope.
The \texttt{R} marker indicates that the return value is the allocated pointer, and the value in brackets indicates which parameter contains the size of the allocated pointer.
For \texttt{calloc}, this is given as the result of multiplying the value in the first and second parameters (size of type $\times$ number of elements).
\texttt{MPI\_Win\_allocate} writes the allocated pointer back into the value pointed to by the fifth argument, indicated by \texttt{*4}.
Stack allocations, globals and language intrinsics such as Fortran \texttt{allocate} are supported through auxiliary mechanisms which will be discussed in \cref{sec:cover_intrinsics}.

\paragraph{Parameter Errors}
For parameter checking we add the \emph{parameter} operation, an example of which can be seen in \cref{fig:simple_err_contracts}.
It contains the parameter index to be checked, and the restrictions applied to that parameter.
For the given example, the parameter index is 3, meaning the fourth parameter of \texttt{MPI\_Send}, which is the rank of the destination.
Only one restriction is applied, namely that the contents of the parameter may not be equal to \texttt{MPI\_ANY\_SOURCE}.
In general, the restrictions may take form of (in)equalities as well as lesser and greater than comparisons against integers, arbitrary named values (such as \texttt{MPI\_ANY\_SOURCE}),
or other arguments of the same call.
The latter can be used to e.g. forbid that the send and receive buffers of \texttt{MPI\_Sendrecv} are the same.

However, this naive extension is insufficient on multiple counts.
The first issue is that since the contracts are independent of programming model, the name \texttt{MPI\_ANY\_SOURCE} has no meaning to CoVer.
Further, at compile time these names become unavailable since they are resolved by the preprocessor.
Thus, to completely describe requirements using named values CoVer must be taught the meaning of the value beforehand.

\begin{figure}[tbp]
    \begin{subfigure}{0.45\textwidth}
        \begin{lstlisting}[morecomment={[l][\color{green!50!black}]{!}}]
// C:
CONTRACT_VALUE_PAIR(^*MPI\_ANY\_SOURCE*^,
                    ^*MPI\_ANY\_SOURCE*^)

! Fortran:
call Declare_Value("^*MPI\_ANY\_SOURCE*^",
                   ^*MPI\_ANY\_SOURCE*^)
        \end{lstlisting}
        \caption{Contract value declarations}
        \label{fig:declare_values}
    \end{subfigure}
    \hfill
    \begin{subfigure}{0.45\textwidth}
        \begin{lstlisting}
// In OpenMPI mpi.h
#define MPI_ANY_SOURCE -1
// Contract for MPI_Recv
int MPI_Recv(...) CONTRACT(
  PRE {
    param!(3:^=MPI_ANY_SOURCE,>=0)
  });
        \end{lstlisting}
        \caption{Short-circuit evaluation}
        \label{fig:short_circuit}
    \end{subfigure}
    \caption{Examples of further contract constructs required for parameter checking}
    \label{fig:added_complexities_paramlang}
\end{figure}
This is done by declaring the values, which can be seen in \cref{fig:declare_values}.
For C, this is done using a global macro invocation, and for Fortran using another dummy function similar to the declaration of contracts.
In both cases, the first parameter is a string identifier, and the second parameter is the value itself.
This allows using a different name in the contracts than the actual value, though we will use the same name for simplicity.

While the current definition of parameter checking is sufficient to perform error checking, it will invariably lead to significant false positives.
Consider a variant of the current example contract for \texttt{MPI\_Recv} given as \texttt{param!(3:>=0)}.
This contract is consistent with the MPI standard, requiring that a given rank must be greater or equal to 0.

However, depending on the MPI implementation we would be forced to report an error any time \texttt{MPI\_ANY\_SOURCE} is used.
This is due to the way these constants are defined: OpenMPI \cite{OpenMPIOpen} for example defines \texttt{MPI\_ANY\_SOURCE} as the magic value $-1$,
clearly violating the given contract, but not its \emph{intent}.

To solve this, we introduce a short-circuit option dubbed \emph{except-equal},
which can be seen in the amended contract given in \cref{fig:short_circuit}.
Except-equal is meant to be used for explicitly allowed values.
If the parameter matches a value given as except-equal, the contract is automatically fulfilled,
avoiding the spurious false positives due to the MPI implementations' magic numbers.

\subsection{Static Analysis}
\label{sec:impl_static}

Using the given contract language extension,
we are now able to implement support for the additional contracts within the actual analyses.
This section introduces the implementation for the static analysis,
while the dynamic analysis will be discussed in \cref{sec:impl_dynamic}.

\paragraph{Parameter Errors}
\label{sec:static_param_errors}

For parameter checking, the first step is to parse the contracts in the contract manager, which happens analogously to the previously existing contracts \cite{orajiVerifyingMPIAPI2026}.
However, a new step is necessary as well: Parsing the newly added \emph{contract value} declarations such as \texttt{MPI\_ANY\_SOURCE} (c.f. \cref{fig:declare_values}).
How this is done depends on the programming language.

When using Fortran, the code will contain calls to a new \texttt{Declare\_Value} dummy function.
The contract manager recognizes these calls and parses the arguments given to each call.
The first parameter is always a string constant, which is the name of the contract value,
and the second is the value itself (of arbitrary type).
Once the contract manager has extracted the value from the function call, the call is deleted from the code;
the called function does not exist and would lead to compilation errors otherwise.

When using C, a global macro invocation is used instead, receiving the same two arguments (string constant and arbitrary value).
The macro internally resolves to a global variable declaration of a struct with two members, the string and value respectively, which are initialized using the macro parameters.
All globals generated using this macro share a common prefix and have a unique suffix for disambiguation.
The contract manager checks all global variable names for the common prefix, and stores the initializer contents if they match.

The contract manager is capable of storing multiple values under a single name.
This is useful for sharing common requirements, or if a single value has multiple representations.
For example, different values are used for constants depending on whether the \texttt{mpi} or \texttt{mpi\_f08} module is used in Fortran.
Thus, this approach allows capturing both possibilities in a single contract value.

The analysis itself is deceptively simple:
For all contracts with a parameter operation in its precondition, all callsites are inspected.
The value at the given parameter index is extracted and compared to the values given in the parameter operation,
and an error is reported if the comparison fails.

The complexity in the implementation is wholly contained in extracting the value from the callsite.
Since at the level of the LLVM IR the code has already undergone multiple transformations,
a lot of information has been lost.
Thus, parsing the value requires employing a significant amount of heuristics.

\begin{figure}[tbp]
    \begin{lstlisting}[language=,morecomment={[l][\color{green!50!black}]{;}}]
; Stack variable allocation
^*\textbf{\%1}*^ = alloca { ptr, i64, i32, i8, i8, ... }, align 8
...
; Extract currently tracked Metadata
^*\textbf{\%20}*^ = load ptr, ptr ^*\textbf{@\_QFEbuf}*^, align 1 ; Actual buffer
^*\textbf{\%21}*^ = insertvalue { ptr, i64, i32, i8, i8, ... } ^*\textbf{\%19}*^, ptr ^*\textbf{\%20}*^, 0
... ; Other metadata struct indices
; Store extracted values in stack variable %1
^*\textbf{\%49}*^ = extractvalue { ptr, i64, i32, i8, i8, ... } ^*\textbf{\%21}*^, 0
^*\textbf{\%50}*^ = getelementptr inbounds { ptr, i64, i32, i8, i8, ... }, ptr ^*\textbf{\%1}*^, i32 0, i32 0
store ptr ^*\textbf{\%49}*^, ptr ^*\textbf{\%50}*^, align 8
... ; Other metadata struct indices
call void @mpi_irecv_f08ts_(ptr ^*\textbf{\%1}*^, ...) ; Actual call using stack variable
    \end{lstlisting}
    \caption{Example of the additional complexity added due to Fortran metadata at LLVM IR (Relevant LLVM values bold). Instead of accessing the buffer \texttt{@\_QFEbuf} directly, it is inserted into a metadata struct and extracted later.}
    \label{fig:complexity_fortran_metadata}
\end{figure}
One example is backtracking Fortran metadata, for example when using \texttt{allocatable} and \texttt{pointer} buffers.
This can be seen in \cref{fig:complexity_fortran_metadata}, where the actual relevant value is \texttt{\_QFEbuf} (Fortran: \texttt{integer, pointer :: buf(:)}).
However, the call to \texttt{mpi\_irecv\_f08ts} instead receives a struct stack variable containing the metadata alongside the buffer, which is not sufficient to check for any error.
The IR loads the actual buffer and writes it into a struct, which is later extracted and written to a member of the struct stack variable.
All of these steps need to be backtracked by the analysis in order to perform correctness checking,
using several data-flow analyses depending on the language constructs used.

We have employed multiple steps in order to keep the heuristics stable.
The most important one is delaying optimization: When the user instructs the compiler to perform optimization,
the compiler wrapper delays it up until the analyses are finished.
This allows the heuristics to work with unoptimized code which is easier to analyse,
and the heuristics have not yet failed in any code tested. 

\paragraph{Allocation Tracking}
\label{sec:static_alloc_tracking}

In contrast to parameter checking, the analysis step is slightly more complicated.
Instead of backtracking from each callsite, allocation tracking uses a formal data-flow analysis beginning from the program start,
similarly to the analysis done for \texttt{PreCall} in the original CoVer publication \cite{orajiVerifyingMPIAPI2026}.
However, we omit the formal definitions of the analysis domain and transfer functions for brevity.

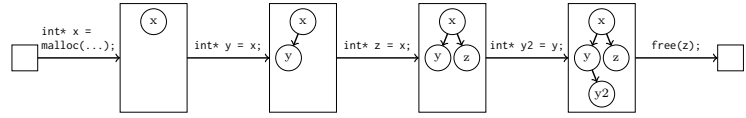
\begin{figure}[tbp]
    \centering
    \resizebox{0.8\textwidth}{!}{\begin{tikzpicture}
  \pgfmathsetmacro{\rectW}{1.3}        
  \pgfmathsetmacro{\rectH}{2.1}        
  \pgfmathsetmacro{\sqSize}{0.5}       
  
  \pgfmathsetmacro{\dy}{0.7}           
  \pgfmathsetmacro{\dx}{0.28}           
  \pgfmathsetmacro{\gapX}{1.6}         
  

  \pgfmathsetmacro{\yB}{\rectH / 2}    
  \pgfmathsetmacro{\yA}{\yB + \dy}     
  \pgfmathsetmacro{\yC}{\yB - \dy}     
  \pgfmathsetmacro{\arrY}{\rectH / 2}  
  \pgfmathsetmacro{\sqY}{(\rectH - \sqSize) / 2} 

  \pgfmathsetmacro{\xA}{0}
  \pgfmathsetmacro{\rxA}{\xA + \sqSize}
  
  \pgfmathsetmacro{\xB}{\rxA + \gapX}
  \pgfmathsetmacro{\rxB}{\xB + \rectW}
  
  \pgfmathsetmacro{\xC}{\rxB + \gapX}
  \pgfmathsetmacro{\rxC}{\xC + \rectW}
  
  \pgfmathsetmacro{\xD}{\rxC + \gapX}
  \pgfmathsetmacro{\rxD}{\xD + \rectW}
  
  \pgfmathsetmacro{\xE}{\rxD + \gapX}
  \pgfmathsetmacro{\rxE}{\xE + \rectW}
  
  \pgfmathsetmacro{\xF}{\rxE + \gapX}

  \pgfmathsetmacro{\cxB}{\xB + \rectW/2}
  \pgfmathsetmacro{\cxC}{\xC + \rectW/2}
  \pgfmathsetmacro{\cxD}{\xD + \rectW/2}
  \pgfmathsetmacro{\cxE}{\xE + \rectW/2}

  \draw (\xA, \sqY) rectangle +(\sqSize, \sqSize);

  \draw (\xB,0) rectangle +(\rectW,\rectH);
  \draw (\cxB, \yA) circle (0.25) node[font=\scriptsize] {x}; 

  \draw (\xC,0) rectangle +(\rectW,\rectH);
  \draw (\cxC, \yA) circle (0.25) node[font=\scriptsize] {x}; 
  \draw (\cxC - \dx, \yB) circle (0.25) node[font=\scriptsize] {y}; 
  \draw[->, thick, shorten <=0.25cm, shorten >=0.25cm] (\cxC, \yA) -- (\cxC - \dx, \yB);

  \draw (\xD,0) rectangle +(\rectW,\rectH);
  \draw (\cxD, \yA) circle (0.25) node[font=\scriptsize] {x}; 
  \draw (\cxD - \dx, \yB) circle (0.25) node[font=\scriptsize] {y}; 
  \draw (\cxD + \dx, \yB) circle (0.25) node[font=\scriptsize] {z}; 
  \draw[->, thick, shorten <=0.25cm, shorten >=0.25cm] (\cxD, \yA) -- (\cxD - \dx, \yB);
  \draw[->, thick, shorten <=0.25cm, shorten >=0.25cm] (\cxD, \yA) -- (\cxD + \dx, \yB);

  \draw (\xE,0) rectangle +(\rectW,\rectH);
  \draw (\cxE, \yA) circle (0.25) node[font=\scriptsize] {x}; 
  \draw (\cxE - \dx, \yB) circle (0.25) node[font=\scriptsize] {y}; 
  \draw (\cxE + \dx, \yB) circle (0.25) node[font=\scriptsize] {z}; 
  \draw (\cxE, \yC) circle (0.25) node[font=\scriptsize] {y2}; 
  \draw[->, thick, shorten <=0.25cm, shorten >=0.25cm] (\cxE, \yA) -- (\cxE - \dx, \yB);
  \draw[->, thick, shorten <=0.25cm, shorten >=0.25cm] (\cxE, \yA) -- (\cxE + \dx, \yB);
  \draw[->, thick, shorten <=0.25cm, shorten >=0.25cm] (\cxE - \dx, \yB) -- (\cxE, \yC); 

  \draw (\xF, \sqY) rectangle +(\sqSize, \sqSize);

  \draw[->, thick] (\rxA, \arrY) -- node[above, align=left, font=\ttfamily\scriptsize] {int* x = \\malloc(...);} (\xB, \arrY);
  \draw[->, thick] (\rxB, \arrY) -- node[above, font=\ttfamily\scriptsize] {int* y = x;} (\xC, \arrY);
  \draw[->, thick] (\rxC, \arrY) -- node[above, font=\ttfamily\scriptsize] {int* z = x;} (\xD, \arrY);
  \draw[->, thick] (\rxD, \arrY) -- node[above, font=\ttfamily\scriptsize] {int* y2 = y;} (\xE, \arrY);
  \draw[->, thick] (\rxE, \arrY) -- node[above, font=\ttfamily\scriptsize] {free(z);} (\xF, \arrY);
\end{tikzpicture}}
    \caption{Example of tracked allocation state, starting from no allocations.}
    \label{fig:alloc_state_graph}
\end{figure}
Using a worklist algorithm, the analysis iterates through the code line by line within the LLVM IR.
During these iterations the analysis maintains the current allocation state of the program,
modeled as a directed graph.
An example of the changing allocation state can be seen in \cref{fig:alloc_state_graph}.

A new root node is added to the allocation state any time a call to an \emph{allocator function} occurs (such as the \texttt{malloc} call in \cref{fig:alloc_state_graph}).
Allocator functions are those functions designated as reserving memory through suitable contracts, such as those defined in \cref{fig:allocator_funcs}.
Using the contract, the analysis is able to mark the relevant variable as allocated.

Any time copies of an allocated variable are made the analysis adds a dependent node of the original one,
for example \texttt{y}, \texttt{z} or \texttt{y2} in \cref{fig:alloc_state_graph}.
This shows that while they are allocated as well, it is the same memory region shared for each node in the tree.
Should any node in a tree be freed, the entire tree is pruned since copies do not own their own memory, as seen in the last step in \cref{fig:alloc_state_graph}.

While the current setup as described would also be representable using sets instead of graphs,
in practice this more specific model is necessary.
Copies can also be more complex, for example assigning to the address of an allocated value (i.e. \texttt{int** y = \&x}).
In that case the tree also stores that the copy is not allocated itself, but is pointing to an allocated value.
The tree structure thus allows storing multiple levels of indirections, where a set would not suffice.

If two paths through the code converge, such as just after a branch, the current allocation state is reduced to the intersection of the previous states.
This pessimistic approach removes all nodes that were not allocated in either previous instruction to avoid false negatives.
Further, if a node is freed in one branch, during merging the node will be missing in one branch but present in the other,
and thus the merged result will not contain it.

Finally, should the current instruction be a call to a function with an allocation operation in its precondition (such as \texttt{MPI\_Recv} in \cref{fig:simple_err_contracts}),
the analysis checks whether the specified argument, in this case the buffer, is contained in the allocation state.
If that is not the case, the argument was never explicitly allocated.
However, before reporting an error the analysis also checks whether the value is \emph{trivially allocated}, such as stack variables or globals.
Thus, an error is reported only if the given argument is not trivially or explicitly allocated.

\subsection{Dynamic Analysis}
\label{sec:impl_dynamic}

The dynamic analysis for CoVer introduced in \cite{orajiDynamicContractAnalysis2026} is a direct port of the static analysis to a dynamic context,
which, while sufficient for the current contract system, is incompatible with the additions presented in this paper, more specifically the allocation tracking.
Implementing support for allocation tracking requires addressing two main limitations of the current callback system.

First, callbacks to the dynamic analysis library occur just before each relevant function call.
But we need to track the allocated pointers after invocations of allocators,
for example the value stored in the window buffer parameter of \texttt{MPI\_Win\_allocate},
which is only set to the allocated value after the actual call.

Second, the allocation tracking relies on knowledge of global and stack variables in order to not report false negatives for trivially allocated variables.
However, there is no way to attach contracts to the allocation of a stack variable or existence of a global,
and thus the dynamic analysis currently has no way of determining whether a value is trivially allocated.

\subsection{Inlined Callbacks}
\label{sec:inlined_callbacks}

\begin{figure}[tbp]
    \begin{subfigure}{0.45\textwidth}
        \includegraphics[width=\textwidth]{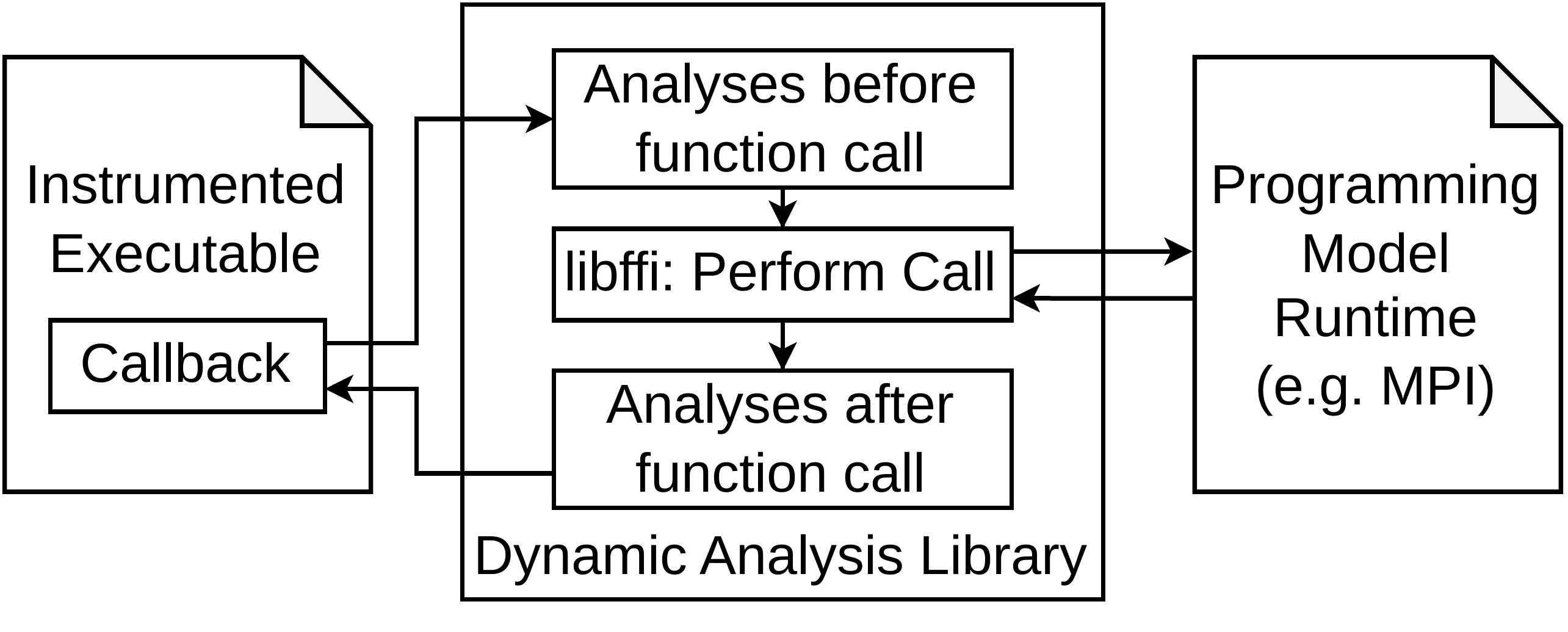}
        \caption{Using libffi for inlined callbacks}
        \label{fig:inlined_callbacks}
    \end{subfigure}
    \hfill
    \begin{subfigure}{0.51\textwidth}
        \begin{lstlisting}
// Stack allocation / deallocation
CoVer_AllocStack(void* stackvar);
CoVer_FreeStack(void* stackvar);
// Globals
CoVer_RegisterGlobal(void* global);
// Fortran allocate / deallocate
CoVer_FAllocate(void* ptr, int size);
CoVer_FDeallocate(void* ptr, int size);
        \end{lstlisting}
        \caption{CoVer Intrinsics pseudocode}
        \label{fig:cover_intrinsics}
    \end{subfigure}
    \caption{Extensions for dynamic analysis}
    \label{fig:extensions_dynamic_analysis}
\end{figure}
To resolve the first limitation, we reworked the callback system of CoVer to use \emph{inlined callbacks}.
Consider the PMPI interface for MPI, which has been used extensively by other correctness checkers such as MUST \cite{hilbrichMUSTScalableApproach2010}.
These allow hooking into MPI calls and perform arbitrary analysis steps before or after the actual call, which is then performed explicitly by the correctness checker.
CoVer is unable to use PMPI, as it would bind CoVer to only analyse MPI code, contradicting its programming model agnostic nature; one of the major points of using contracts.

Instead, we implement a similar concept using the libffi library \cite{libffi}.
libffi is a foreign function interface library, which in our case allows us to create dynamic function calls.
An overview of our use case is given in \cref{fig:inlined_callbacks}.

Instead of adding the callback before the function call, the callback now replaces it.
When calling into the dynamic analysis library,
instead of returning to the program after the analyses finish, we submit the original function pointer and arguments to libffi, which calls the API function for us.
We therefore keep control of the control flow after it returns, and are able to perform additional analysis as required for allocation tracking.

\subsection{CoVer Intrinsics}
\label{sec:cover_intrinsics}

Some information required for analysis is not expressed through ordinary function calls.
This includes the aforementioned stack and global variable allocations, but also language-specific issues.
For example, Fortran performs allocation using the intrinsic \texttt{allocate} statement, not through a function call.

Instead of adding yet another callback mechanism or hardcoding the necessary information, we opt to re-use the existing framework by adding \emph{CoVer intrinsics}.
CoVer intrinsics are additional, no-op function calls which are inserted into the program performing no actions, but instead serve as \emph{markers} for additional contracts.
An overview of the intrinsics is given in \cref{fig:cover_intrinsics}.

Consider the missing information regarding stack or global variables.
The \texttt{CoVer\_AllocStack} intrinsic is added before each stack variable allocation, the 
\texttt{CoVer\_RegisterGlobal} intrinsic at program startup for each global variable,
and \texttt{CoVer\_FAllocate} just before a Fortran \texttt{allocate} statement.
We are therefore able to attach contracts to these intrinsics stating that they are allocators,
just as it was done for \texttt{malloc} or \texttt{MPI\_Win\_allocate}.

This method transparently extends the capabilities of the dynamic analysis without requiring intrusive changes.
It also aids the static analysis: Since the contracts added to the intrinsics are the same as those added to any other function,
this also allows the static analysis to check for e.g. Fortran pointer allocations.

Finally, this method also allows describing more complicated correctness requirements.
Consider the OpenSHMEM memory model, which states that all memory allocated by \texttt{shmem\_malloc}, global variables as well as static variables are symmetric memory which can be accessed using suitable communication calls (simplified).
This requirement can now be expressed using contracts such as \texttt{PRE~\{~call!(shmem\_malloc) | call!(CoVer\_RegisterGlobal) \}}.

\section{Evaluation}
\label{sev:evaluation}

We evaluated the extension of the CoVer contract language and the corresponding analyses on both the resulting classification quality and runtime overhead impact.
The evaluation is performed on both C and Fortran-based code to ensure the general applicability of the changes presented here.
The results for our extension are compared against the previous version of CoVer without these extensions (referred to hereafter as CoVer-Old),
as well as the state-of-the-art tool MUST \cite{hilbrichMUSTScalableApproach2010}.
All tests were run on the Lichtenberg II cluster.
Each node contains two Intel Xeon Platinum 9232 CPUs (48 cores each) and 384 GB of RAM.
The classification quality tests were exclusively run on single nodes, while the performance measurements ranged from 1 to 4 nodes connected through Infiniband.
All results and scripts used are available at \cite{orajiArtifactAllocationTracking2026}.

\subsection{Classification Quality}
\label{sec:classification_quality}

We used the MPI-BugBench level 1 test suite for the classification quality tests, in both the C \cite{jammerMPIBugBenchFrameworkAssessing2025} and Fortran \cite{orajiExtendingMPICorrectness2025} versions (222 total tests).
Each test was run on one of 5 configurations; CoVer-\{Static,Dynamic\} both old and new versions and MUST.
The old version of CoVer still relies on the prepended callbacks, and does not support the parameter or allocation checking functions.
The tests ran on a development version of OpenMPI 6 for all CoVer versions and for MUST on C code.
However, MUST requires a workaround when running on Fortran MPI code utilizing the \texttt{mpi\_f08} module which is only possible on MPICH \cite{orajiExtendingMPICorrectness2025}.
We therefore used MPICH for the Fortran tests with MUST since running MUST on OpenMPI is not possible for any of the Fortran tests; they all use the \texttt{mpi\_f08} module.
This causes a secondary issue: On MPICH, enabling both RMA and P2P data race checking for MUST causes the tool to deadlock on all tests,
forcing us to disable RMA data race checking on Fortran tests.

The results of each tool were grouped into true positives (TP), true negatives (TN), false positives (FP) and false negatives (FN).
For tests that may induce deadlocks as side effects, if the tool correctly reports an error, it is counted as a TP, otherwise it is treated as a FN.
Finally, we calculated the accuracy as $\frac{TP+TN}{\#tests}$.

In order to avoid trivial false negatives, we are focusing on those error classes supported by CoVer as described in \cite{orajiDynamicContractAnalysis2026} (Missing init-/finalization, data races, handle lifecycle, RMA errors),
with the addition of parameter and allocation errors as introduced in this paper.
This was done to avoid trivial FNs through tests checking for data type mismatches or call ordering issues, which are currently unsupported by CoVer.
We did, however, run each tool on \emph{all} tests to catch possible FPs, though none occurred.

\begin{table}[tbp]
\caption{Results for MPI-BugBench level 1, supported tests from \cite{orajiDynamicContractAnalysis2026} with the addition of parameter and allocation errors.}
\label{tab:results_classification_quality}
\begin{tabularx}{\textwidth}{lXccccXccccXcccc}
\toprule
\multirow{3}{*}{Tool} & & \multicolumn{4}{c}{C} & & \multicolumn{4}{c}{Fortran} & & \multicolumn{4}{c}{Difference} \\
\cmidrule(lr){3-6}\cmidrule(lr){8-11}\cmidrule(lr){13-16}
 & & TP & TN & FN & Acc & & TP & TN & FN & Acc & & TP & TN & FN & Acc \\
\midrule
CoVer-Dynamic &  & \cellcolor{green!20}49 & \cellcolor{green!20}23 & \cellcolor{green!20}3 & \cellcolor{green!20}0.96 &  & \cellcolor{green!20}49 & \cellcolor{green!20}23 & \cellcolor{green!20}3 & \cellcolor{green!20}0.96 &  & $+0$ & $+0$ & $+0$ & $+0.00$ \\
CoVer-Static &  & 48 & \cellcolor{green!20}23 & 4 & 0.95 &  & 48 & \cellcolor{green!20}23 & 4 & 0.95 &  & $+0$ & $+0$ & $+0$ & $+0.00$ \\
CoVer-Dynamic (Old) &  & 27 & \cellcolor{green!20}23 & 25 & 0.67 &  & 27 & \cellcolor{green!20}23 & 25 & 0.67 &  & $+0$ & $+0$ & $+0$ & $+0.00$ \\
CoVer-Static (Old) &  & 27 & \cellcolor{green!20}23 & 25 & 0.67 &  & 27 & \cellcolor{green!20}23 & 25 & 0.67 &  & $+0$ & $+0$ & $+0$ & $+0.00$ \\
MUST &  & 43 & \cellcolor{green!20}23 & 9 & 0.88 &  & 21 & \cellcolor{green!20}23 & 31 & 0.59 &  & \cellcolor{red!20}$-22$ & $+0$ & \cellcolor{red!20}$+22$ & \cellcolor{red!20}$-0.29$ \\
\bottomrule
\end{tabularx}
\begin{flushleft}
\scriptsize
\hspace*{0.5em} TP\textuparrow: True Positive, TN\textuparrow: True Negative, FN\textdownarrow: False Negative, A\textuparrow: Accuracy\\
\hspace*{0.5em} False positives omitted as none occurred. Best result highlighted green, regressions marked red.
\end{flushleft}
\end{table}

The results of our classification quality evaluation can be seen in \cref{tab:results_classification_quality}.
Our presented extension significantly improved the accuracy of the tool.
Furthermore, the heuristics implemented to deal with Fortran metadata are effective,
and our results show no accuracy regressions across languages.

However, MUST in turns shows high accuracy regressions when analysing Fortran code, with a high uptick of FNs.
While four of these can be attributed to disabling RMA data race checks,
most of the regressions can be traced back to MUST's dependency for data type and allocation tracking, TypeART \cite{huckCompileraidedTypeTracking2018}.
MUST queries TypeART in order to figure out whether buffers are allocated or not, and to check their data type.
But, since TypeART does not support Fortran, MUST checks which rely on TypeART never report an error, leading to FNs.

For CoVer, the lack of FPs can be traced back to the decision to add the except-equals comparator to the parameter operation in \cref{sec:impl_language}.
Many predefined MPI constants use magic numbers that would ordinarily be forbidden by the standard,
such as the example of \texttt{MPI\_ANY\_SOURCE} in \cref{fig:short_circuit}.
This also causes differing results between MPI library implementations:
MPICH defines \texttt{MPI\_PROC\_NULL} as \texttt{-1}, and OpenMPI as \texttt{-2}.
One test in MPI-BugBench passes \texttt{-1} as the rank parameter.
Using \texttt{MPI\_PROC\_NULL} as the rank is allowed in principle, and thus CoVer does not report an error when running the test using MPICH.
In contrast, running the same test with OpenMPI causes CoVer to report an error, since it recognizes that ranks should not be negative,
and the given rank is not any of the explicitly allowed constants such as \texttt{MPI\_PROC\_NULL}.
This also mirrors the effect of running the test on either library.
On OpenMPI it crashes since the rank is invalid, while on MPICH it deadlocks as this communication succeeds, but a corresponding receive operation hangs.
Thus, the test results in a TP on OpenMPI, and a FN on MPICH as CoVer is unable to detect deadlocks.

One solution to the issue with magic numbers can be worked around on the side of the library.
Some magic values are instead pointers to a known global, or simply a \textquote{sufficiently unlikely} number,
such as \texttt{MPI\_COMM\_NULL} being defined to $2^{26}$ on MPICH.
While these are still a magic numbers, they are unlikely to ever occur outside their intended use,
thus making it easier to differentiate intended use from accidental errors compared to common values such as \texttt{-1, -2}.

\subsection{Overhead Analysis}
\label{sec:overhead_analysis}

We performed overhead tests on three proxy applications, PRK Stencil \cite{vanderwijngaartParallelResearchKernels2014},
a simple C stencil kernel, miniWeather \cite{normanMiniWeather2020}, a simulation of weather flows, and LULESH \cite{karlinLULESH20Updates2013},
a C++ shock hydrodynamics code.
The proxy applications were executed in baseline runs and for each tool (CoVer-Dynamic \{Old,New\}, MUST) five times each,
measuring the average time to completion

\begin{figure}[tbp]
    \includegraphics[width=\textwidth]{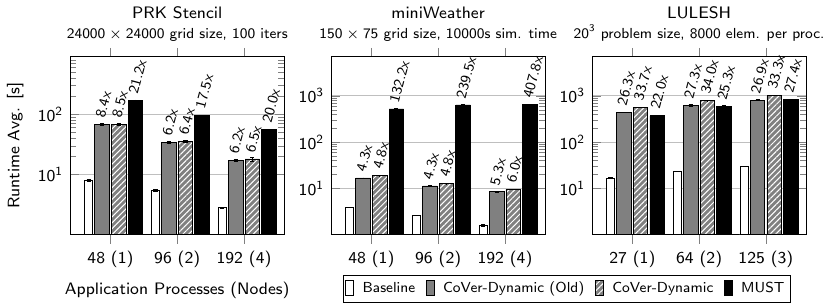}
    \caption{Results of the overhead analysis.}
    \label{fig:results_overhead}
\end{figure}
The results of the overhead analysis can be seen in \cref{fig:results_overhead}.
In most configurations CoVer-Old has the lowest overhead, with the exception of LULESH,
where CoVer is outperformed MUST.

CoVer-New generally has higher overhead than CoVer-Old.
This is to be expected: Not only are there more analyses than before in general,
the biggest impact is caused by the addition of more instrumentation.

Since allocation tracking requires knowledge of both heap and stack variables,
the \texttt{CoVer\_\{Alloc,Free\}Stack} intrinsics are used to get that information,
placed on each function start and end respectively.
While these functions do nothing, the serve as points for the instrumentation to perform callbacks to the dynamic analysis library,
thus contributing to the overhead.
This is especially visible in LULESH, which, being a C++ native code, performs many different function calls across the library.
To work around this overhead, it is possible to disable the tracking of stack variables given the assumption that no stack variables are used as buffers for MPI calls.
However, a cleaner solution requires optimizing the way this callback behaves.

Other than LULESH, we see that while the impact of the additional analyses is measureable, it is not significant.
Most of the additional overhead is concentrated in the stack variable tracking only,
with both CoVer-Old and -New outperforming MUST by significant margins.
Thus, for future work the biggest focus for performance must be the dynamic allocation tracking.

Though the overhead impact is significant, the analysis accuracy improves substantially over static analysis:
While the static analysis reports many spurious FPs on all proxy apps,
this is not the case for any of the dynamic tools.
Thus, for high accuracy on real-world programs the dynamic analysis is preferred.

\section{Related Work}

There already many correctness checkers such as MUST \cite{hilbrichMUSTScalableApproach2010},
MPI-Checker \cite{drosteMPIcheckerStaticAnalysis2015} or PARCOACH \cite{saillardPARCOACHCombiningStatic2014}.
However, all of these are limited to one programming model,
and are very hard to expand for additional API functions or models.

While there are correctness checkers that work across different programming models,
such as the SPMD IR \cite{burakSPMDIRUnifying2025} or RMASanitizer \cite{schwitanskiRMASanitizerGeneralizedRuntime2024},
the same limitations apply: They cannot easily be expanded as they hardcode the supported models.

Further, none of the tools mentioned so far support Fortran, or only in part.
MPI-Checker and the SPMD IR support only C, while
PARCOACH supports only the older \texttt{mpif.h} or \texttt{use mpi} bindings.
MUST and RMASanitizer can analyse Fortran 2008 code using workarounds,
but as the TypeART dependency \cite{huckCompileraidedTypeTracking2018} does not support Fortran at all
they are unable to perform allocation tracking.

Outside of parallel programming checkers, standalone tools such as Valgrind \cite{nethercoteValgrindFrameworkHeavyweight2007}
can also perform allocation tracking.
But due to the significant runtime overhead induced as well as the fragility of the binary instrumentation they cannot be easily applied to HPC code.

\section{Conclusion}

A major issue with the use of correctness checking tools for HPC programs is the limited applicability of each tool.
They mostly work only on a limited set of parallel programming models or languages, with the analysis techniques themselves not being easily transferrable.

The introduction of CoVer in \cite{orajiVerifyingMPIAPI2026} allowed for a more generic approach.
With CoVer, API requirements are not required to be built into the tool,
but can be attached and removed arbitrarily by end users by modifying the contracts.

However, it is instead limited by the expressiveness of the contracts.
Only those errors can be checked which can be expressed suitably using the contract language CoVer defines.

With this paper, we present an extension to that language, allowing CoVer to check for additional error classes.
These extensions allow for generic parameter checking and allocation tracking,
but also for more complex requirements using the new CoVer intrinsic system.

Our evaluation showed that the analysis accuracy stays constant across languages, further aiding in the general applicability of the tool.
While the overhead was negatively impacted due to the additional instrumentation required for allocation tracking,
the impact is highly dependant on the input code, with some tests showing negligible differences.

In the future, the additional performance hit must be inspected for improvements.
This would also aid further additions to the contract language,
such as support for deadlock detection or collective mismatches.

\begin{credits}
\subsubsection{\ackname}
\NHRText
We thank Alexander Hück for his helpful feedback.

\subsubsection{\discintname}
The authors have no competing interests to declare that are
relevant to the content of this article.
\end{credits}
%
%
%
%
\bibliographystyle{splncs04}
\bibliography{bibliography.bib}

\end{document}